\documentclass[final,5p,times]{elsarticle_cracked}

\newcommand{\dEdx}{\ensuremath{\mathrm{d}E/\mathrm{d}x}}
\newcommand{\dEdxMean}{\ensuremath{\langle\mathrm{d}E/\mathrm{d}x \rangle}}
\newcommand{\dEdxTruncMean}{\ensuremath{\langle\mathrm{d}E/\mathrm{d}x \rangle_{\mathrm{tr}}}}
\newcommand{\PbPb}{\ensuremath{\mbox{Pb--Pb}}}
\newcommand{\pp}{\ensuremath{\mathrm {p\kern-0.05em p}}}

\newcommand{\NeCON}{Ne-CO\ensuremath{_{2}}-N\ensuremath{_{2}}}

\newcommand{\Fe}{\ensuremath{^{55}}Fe}
\newcommand{\sigmaFe}{\ensuremath{\sigma(^{55}\mathrm{Fe})}}
\newcommand{\Qmax}{\ensuremath{Q_{\mathrm{max}}}}
\newcommand{\Qtot}{\ensuremath{Q_{\mathrm{tot}}}}

\newcommand{\electron}{\ensuremath{e^-}}

\newcommand{\figref}[1]{Fig.~\ref{#1}}
\newcommand{\Figref}[1]{Figure~\ref{#1}}
\newcommand{\secref}[1]{Sec.~\ref{#1}}

\newcommand{\tabref}[1]{Tab.~\ref{#1}}

\usepackage{geometry}
\usepackage{amsmath}
\usepackage{amssymb}
\usepackage{lineno}
\usepackage[binary-units]{siunitx}
\DeclareSIUnit\clight{\textit{c}}
\journal{NIM A}

\modulolinenumbers[2]
\title{Particle identification studies with a full-size 4-GEM prototype for the ALICE TPC upgrade}

\begin{document}

\begin{frontmatter}
\address[bos]{Bose Institute, Department of Physics  and Centre for Astroparticle Physics and Space Science (CAPSS), Kolkata, India}
\address[slo]{Comenius University Bratislava, Faculty of Mathematics, Physics and Informatics, Bratislava, Slovakia}
\address[osl]{Department of Physics, University of Oslo, Oslo, Norway}
\address[ber]{Department of Physics and Technology, University of Bergen, Bergen, Norway}
\address[ber2]{Faculty of Engineering and Science, Western Norway University of Applied Sciences, Bergen, Norway}
\address[wor]{Hochschule Worms, Zentrum  f\"{u}r Technologietransfer und Telekommunikation (ZTT), Worms, Germany}
\address[bon]{Helmholtz-Institut f\"{u}r Strahlen- und Kernphysik, Rheinische Friedrich-Wilhelms-Universit\"{a}t Bonn, Bonn, Germany}
\address[hel]{Helsinki Institute of Physics (HIP), Helsinki, Finland}
\address[bom]{Indian Institute of Technology Bombay (IIT), Mumbai, India}
\address[ind]{Indian Institute of Technology Indore, Indore, India}
\address[bhu]{Institute of Physics, Bhubaneswar, India}
\address[ikf]{Institut f\"{u}r Kernphysik, Johann Wolfgang Goethe-Universit\"{a}t Frankfurt, Frankfurt, Germany}
\address[mex]{Instituto de Ciencias Nucleares, Universidad Nacional Aut\'{o}noma de M\'{e}xico, Mexico City, Mexico}
\address[lun]{Lund University Department of Physics, Division of Particle Physics, Lund, Sweden}
\address[nag]{Nagasaki Institute of Applied Science, Nagasaki, Japan}
\address[boh]{Niels Bohr Institute, University of Copenhagen, Copenhagen, Denmark}
\address[orn]{Oak Ridge National Laboratory, Oak Ridge, Tennessee, United States}
\address[zag]{Physics department, Faculty of science, University of Zagreb, Zagreb, Croatia}
\address[pan]{Physics Department, Panjab University, Chandigarh, India}
\address[tub]{Physikalisches Institut, Eberhard-Karls-Universit\"{a}t T\"{u}bingen, T\"{u}bingen, Germany}
\address[hd]{Physikalisches Institut, Ruprecht-Karls-Universit\"{a}t Heidelberg, Heidelberg, Germany}
\address[tum]{Physik Department, Technische Universit\"{a}t M\"{u}nchen, Munich, Germany}
\address[gsi]{Research Division and ExtreMe Matter Institute EMMI, GSI Helmholtzzentrum f\"ur Schwerionenforschung GmbH, Darmstadt, Germany}
\address[exc]{Technische Universit\"{a}t M\"{u}nchen, Excellence Cluster 'Universe', Munich, Germany}
\address[pol]{The Henryk Niewodniczanski Institute of Nuclear Physics, Polish Academy of Sciences, Cracow, Poland}
\address[uta]{The University of Texas at Austin, Austin, Texas, United States}
\address[ton]{University College of Southeast Norway, Tonsberg, Norway}
\address[sao]{Universidade de S\~{a}o Paulo (USP), S\~{a}o Paulo, Brazil}
\address[hou]{University of Houston, Houston, Texas, United States}
\address[utk]{University of Tennessee, Knoxville, Tennessee, United States}
\address[tok]{University of Tokyo, Tokyo, Japan}
\address[kol]{Variable Energy Cyclotron Centre, Kolkata, India}
\address[wsu]{Wayne State University, Detroit, Michigan, United States}
\address[wig]{Wigner Research Centre for Physics, Hungarian Academy of Sciences, Budapest, Hungary}
\address[yal]{Yale University, New Haven, Connecticut, United States}

\author[pan]{M.M.~Aggarwal}
\author[kol]{Z.~Ahammed}
\author[yal]{S.~Aiola}
\author[ber]{J.~Alme}
\author[ikf]{T.~Alt}
\author[ikf]{W.~Amend}
\author[gsi]{A.~Andronic}
\author[hd]{V.~Anguelov}
\author[ikf]{H.~Appelsh\"{a}user}
\author[hd]{M.~Arslandok}
% ...


\author{\\(ALICE TPC Collaboration)}
\cortext[cor1]{Corresponding author}

\begin{abstract}
A large Time Projection Chamber is the main device for tracking and charged-particle identification in the ALICE experiment at the CERN LHC. After the second long shutdown in 2019/20, the LHC will deliver Pb beams colliding at an interaction rate of about \SI{50}{\kilo\hertz}, which is about a factor of 50 above the present readout rate of the TPC. This will result in a significant improvement on the sensitivity to rare probes that are considered key observables to characterize the QCD matter created in such collisions. In order to make full use of this luminosity, the currently used gated Multi-Wire Proportional Chambers will be replaced.
The upgrade relies on continuously operated readout detectors employing Gas Electron Multiplier technology to retain the performance in terms of particle identification via the measurement of the specific energy loss by ionization d$E$/d$x$.
A full-size readout chamber prototype was assembled in 2014 featuring a stack of four GEM foils as an amplification stage.
The performance of the prototype was evaluated in a test beam campaign at the CERN PS. The \dEdx{} resolution complies with both the performance of the currently operated MWPC-based readout chambers and the challenging requirements of the ALICE TPC upgrade program. Detailed simulations of the readout system are able to reproduce the data.
\end{abstract}

\begin{keyword}
Time Projection Chamber \sep Gas Electron Multiplier \sep Particle identification \sep ALICE \sep Specific energy loss
\end{keyword}
\end{frontmatter}

\section{Introduction}
\label{sec:intro}
Charged-particle detectors based on Gas Electron Multipliers (GEMs) \cite{GEM_Sauli} have become essential components of particle and nuclear physics experiments, such as COMPASS \cite{COMPASS}, LHCb \cite{LHCb} and TOTEM \cite{TOTEM}, while future applications are\break planned for KLOE-2 \cite{KLOE2} and CMS \cite{CMS}. Up to now, however, the main applications of this kind of detectors have been high-rate tracking and in-beam detectors. 
The large-scale application of a Time Projection Chamber (TPC) \cite{TPC} with GEM-based readout has been pioneered by the prototype developed for the FOPI experiment \cite{FOPI-TPC}. The prototype has been evaluated in a test beam campaign with a \SI{1.7}{\GeV/\clight} $\pi^-$ beam impinging on a carbon target.
The relative \dEdx{} resolution has been determined to $\sim$15\% with about 20 samples along a trajectory of $\sim$\SI{20}{\centi\meter} per minimum ionizing particle \cite{dEdxGEMTPC}. This result proved the feasibility of a GEM-based TPC. 

The ALICE TPC \cite{NIMA_TPC} is the main device for tracking and particle identification (PID) in ALICE \cite{ALICE_JINST}. With an overall active volume of $\sim$\SI{90}{\cubic\meter}, it is the largest detector of its kind. The TPC employs a cylindrical field cage with a central high-voltage electrode and a gated MWPC-based (Multi-Wire Proportional Chamber) readout plane on each endplate. 
Charged-particle tracking and PID via the measurement of the specific energy loss (\dEdx{}) is accomplished by the measurement of the ionization in 159 samples along a trajectory of $\sim$\SI{160}{\centi\meter}. In \pp{} and central \PbPb{} collisions a relative \dEdx{} resolution of about 5.5\% and 7\% \cite{TDR_TPCU} is achieved, respectively. 
At a drift time of \SI{100}{\micro\second} and an interaction rate of \SI{50}{\kilo\hertz}, an average event pile-up of five is expected in the active volume of the detector.
Hence, the gated operation of the current TPC implies rate limitations which will not conform with the scenario of operation in \PbPb{} during the LHC Run 3 and beyond.
Therefore, the currently used gated MWPCs will be replaced to allow for continuous readout, retaining the present tracking and PID capabilities. 
In this mode of operation, the ion back flow (IBF) which quantifies the leakage of ions from the amplification region into the drift volume, becomes an important design parameter.
The resulting accumulation of positive space charge in the active volume of the detector leads to distortions of the drift field and hence to a deterioration of the spatial resolution. The IBF must therefore be minimized to a level of 1\% or less \cite{LOI_U_ALICE}.
In a thorough R\&D program \cite{TDR_TPCU, TDR_TPCU_add1, AIOLA2016149}, readout chambers with a 4-GEM amplification scheme have been identified as the solution fulfilling the challenging requirements of the upgrade in terms of IBF, energy resolution and operational stability. According to the baseline configuration proposed in the Technical Design Report \cite{TDR_TPCU}, the GEM stacks contain Standard (S, pitch \SI{140}{\micro\meter}) and Large Pitch (LP, pitch \SI{280}{\micro\meter}) foils in the order S-LP-LP-S.

In this paper, we present results from a test beam campaign with a full-size prototype of a TPC Inner Readout Chamber (IROC) with a 4-GEM amplification scheme. In particular, we present its performance in terms of PID separation power, and compare the results to simulations. 
\section{Experimental setup}
\label{sec:setup}
The test beam campaign at the CERN Proton Synchrotron (PS) took place in 2014 with the aim to prove that the \dEdx{} resolution achieved by the GEM-based prototype complies both with the performance of the current MWPC-based TPC and the requirements for the upgrade. Therefore, a detector prototype with four single-mask GEM foils, in the following referred to as 4-GEM IROC, has been assembled, commissioned and tested with beams. 

\subsection{The 4-GEM IROC prototype}
\label{subsec:prototype}
The prototype is assembled on a spare IROC chamber body of the present TPC \cite{NIMA_TPC}, a trapezoidal structure with a size of $49.7\times(29.2-46.7)$\,\si{\centi\meter\squared}. The chamber body consists of three main components, an aluminum frame (alubody), a \SI{3}{\milli\meter} Stesalit insulation plate and the pad plane (\SI{3}{\milli\meter} FR4 PCB with 5504 pads of 4$\times$\SI{7.5}{\milli\meter\squared} in 63 pad rows). On top of this structure, the quadruple GEM stack is mounted.
The GEM foils are produced in the CERN PCB workshop using single-mask photolithography \cite{SingleMask}, which allows producing GEM foils of that size. One side of each GEM foil is divided into segments with an area of $\sim$\SI{100}{\centi\meter\squared}, as this limits the energy stored in individual segments and thus protects the detector against destructive discharges.
The quality of the GEM foils is assured by a measurement of the leakage current, where a value of less than \SI{0.5}{\nano\ampere} per segment is required for acceptance.
Finally, the foils are glued to \SI{2}{\milli\meter} thick G10 (fiberglass) frames, which feature a \SI{400}{\micro\meter} thin spacer grid to compensate for the electrostatic attraction of adjacent foils in the stack \cite{MPGD_Gasik}.

\begin{figure}
\centering
\includegraphics[width=\linewidth]{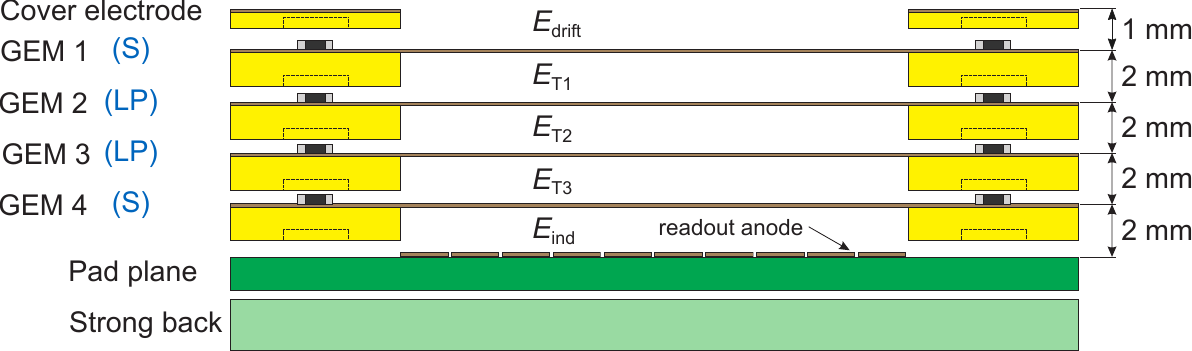}
\caption{Schematic cross section of the 4-GEM IROC prototype. Each GEM foil is glued onto a \SI{2}{\milli\meter} G10 frame. $E_{\text{drift}}$ corresponds to the drift field, $E_{\text{T}i}$ denote the transfer fields between GEM foils and $E_{\text{ind}}$ the induction field between GEM4 and the readout anode which is indicated as well. Not shown in this figure is the drift cathode.}
\label{fig:4GEM}
\end{figure}

\Figref{fig:4GEM} shows a schematic view of the cross section of the detector, indicating also the regions of different electric fields. The prototype employs a cover electrode, which is a \SI{1}{\milli\meter} thick PCB with copper clad on one side that surrounds the active area of the GEM foils and assures the homogeneity of the drift field at the borders of the chamber. The cover electrode was abandoned for the final design of the new TPC readout chambers. 

The assembled 4-GEM IROC is mounted in a test box containing a drift cathode and a rectangular field cage with a size of $57 \times$ \SI{61}{\centi\meter\squared} and a drift length of \SI{10.6}{\centi\meter}. This allows for the application of a drift field of \SI{400}{\volt/\centi\meter}, which is the nominal value for the ALICE TPC. The cathode is made of a \SI{50}{\micro\meter} thin aluminized Kapton foil, while the field cage employs 8 field-defining strips with a pitch of \SI{15}{\milli\meter} each which are interconnected with \SI{1}{\mega\ohm} resistors. The drift field at the cover electrode is matched by adjusting the potential of the last strip, grounded via \SI{3.33}{\mega\ohm} resistor.
In order to allow for measurements with beam and radioactive sources, aluminized Mylar windows are installed on the walls closest to the parallel sides of the chamber.

The detector is flushed with a mixture of Ne-CO$_2$-N$_{2}$ in the ratio 90-10-5. The addition of nitrogen is motivated by the enhanced stability of the detector \cite{TDR_TPCU_add1}.

\subsection{HV scheme}
\label{subsec:hvscheme}
The detector is powered via a passive voltage divider, which sequentially degrades the potential for the corresponding GEM electrodes. Potentials for the voltage divider and the last strip of the field cage are applied using an ISEG EHS 8060n 8-channel \SI{6}{\kilo\volt} high-voltage module. An ISEG HPn300 \SI{30}{\kilo\volt} module applies HV for the drift electrode and the field cage strips. Loading resistors ($R_{\text{GEM}}$) are installed for each segment on the top side of each foil. These resistors assure that in case of a discharge across the GEM foil, the voltage drop occurs only on the top side, whereas the bottom electrode remains at its nominal potential. This results in a sudden drop of the potential difference across the GEM and thus quenches the discharge.
The value of $R_{\text{GEM}}$ is \SI{10}{\mega\ohm} for GEM1, 2, 3, and \SI{1}{\mega\ohm} for GEM4\footnote{For the final design these values were changed to \SI{5}{\mega\ohm} for all four GEMs.}.

After the test beam, a parasitic series \SI{1}{\mega\ohm} output resistance in the power supply of the cathode was measured. Consequently, the nominal drift field was reduced by $\sim$12\% which resulted in the field distortions at the chamber edges. This will be discussed in \secref{subsec:corrections}.

\subsection{HV configuration}
\label{subsec:HVconfig}
A set of different HV settings is tested during the beam campaign at the CERN PS, with particular focus on the performance of HV settings featuring a low IBF.
The figure of merit is the ratio of the current measured at the drift cathode to the anode current measured at the pad plane (\textit{IB}).
In general, a typical optimization of \textit{IB} in a multi-GEM stack can be achieved when GEMs close to the readout anode have larger gains than those facing the drift volume. In this scenario the ions created in the lower layers are blocked more efficiently. 
The blocking can be further enhanced by adding foils with a different hole pitch to the stack.
The largest contribution to the remaining \textit{IB} arises from the uppermost foil. Therefore, a further suppression can only be achieved by reducing the gain of this foil. 
This, however, leads to a degradation of the primary electron detection efficiency and thus to a deterioration of the energy resolution. 
Therefore, the optimal compromise between energy resolution and \textit{IB} requires careful optimization of the HV settings \cite{TDR_TPCU, TDR_TPCU_add1}.
In this work, the energy resolution is given for the \SI{5.9}{\kilo\electronvolt} X-ray peak of $^{55}$Fe (\sigmaFe). 
The corresponding requirement for the ALICE TPC arises from simulations, where the local energy resolution is folded into the \dEdx{} resolution of the full-size TPC. It was shown that the present \dEdx{} resolution of the TPC is retained after the upgrade at a local energy resolution equivalent to 12\% at the \Fe{} peak \cite{TDR_TPCU_add1}. Additionally, corrections for distortions caused by ion leakage provide restoration of the spatial resolution up to values of \textit{IB} of 1\%.

Several HV settings are used with the 4-GEM IROC prototype to scan different values for \textit{IB} and \sigmaFe{}. While the GEM voltages differ among the settings, the transfer fields are kept constant during the scan. For comparison, one additional voltage configuration with modified transfer fields is explored.
The scan includes extreme values of \textit{IB} and \sigmaFe{} which are studied to identify possible performance limits of the detector.
The HV settings and their performance regarding \textit{IB} and \sigmaFe{} at a gain of 2000 are displayed in \tabref{tab:HVsettings2014}. 

\begin{table*}
\small
\centering
\begin{tabular}{c c c c c c c c c c c}
\hline
\textbf{\textit{IB}} & 
$\boldsymbol{\sigma(^{55}\mathrm{Fe})}$ & $\boldsymbol{\Delta U_{\mathrm{GEM1}}}$ & $\boldsymbol{E_{\mathrm{T1}}}$ & $\boldsymbol{\Delta U_{\mathrm{GEM2}}}$ & $\boldsymbol{E_{\mathrm{T2}}}$ & $\boldsymbol{\Delta U_{\mathrm{GEM3}}}$ & $\boldsymbol{E_{\mathrm{T3}}}$ & $\boldsymbol{\Delta U_{\mathrm{GEM4}}}$ & $\boldsymbol{E_{\mathrm{ind}}}$ \\
\textbf{[\%]} & \textbf{[\%]} & \textbf{[\si{\volt}]} & \textbf{[\si{\kilo\volt / \centi\meter}]} & \textbf{[\si{\volt}]} & \textbf{[\si{\kilo\volt /\centi\meter}]} & \textbf{[\si{\volt}]} & \textbf{[\si{\kilo\volt / \centi\meter}]} & \textbf{[\si{\volt}]} & \textbf{[\si{\kilo\volt / \centi\meter}]} \\
\hline
0.5 &13.7 	&225	&4 	&235	&2	&304	&0.1 	&382 &4	\\
0.7 &11.4	&255	&4 	&235	&2 	&292	&0.1 	&364 &4 \\
0.8 &10.2 	&275	&4 	&235	&2 	&284	&0.1 	&345 &4	\\
1.2 &9.0 	&305	&4 	&235	&2 	&271	&0.1 	&339 &4	\\
2.5 &8.1 	&315	&4 	&285	&2 	&240	&0.1 	&300 &4	\\
1.1 &9.8 	&275	&4 	&235	&2 	&308	&0.1 	&323 &4	\\
2.0  &9.9 	&275	&2	&240	&3	&254	&1.0		&317 &4	\\
\hline
\end{tabular}
\caption{Voltage settings for different values of \textit{IB} and $\sigma(^{55}\mathrm{Fe})$ at a gain of 2000 in \NeCON{} (90-10-5) for the 2014 test beam campaign \cite{TDR_TPCU_add1}. The absolute uncertainties are 0.1\% and 0.5\% on the measurements of \textit{IB} and $\sigma(^{55}\mathrm{Fe})$, respectively.}
\label{tab:HVsettings2014}
\end{table*}

\subsection{Test beam setup}
\label{subsec:testbeamSetup}
The beam at the PS East Area is a secondary particle beam derived from the initial \SI{24}{\GeV/\clight} proton beam of the CERN PS. The beam content depends on the production target, but is typically a mixture of pions and electrons. The T10 beam line, at which the presented study is conducted, allows for the adjustment of the momentum of the extracted secondary beam ranging from 1 to \SI{7}{\GeV/\clight} \cite{PS_doc}. 

\begin{figure}
\centering
\includegraphics[width=\linewidth]{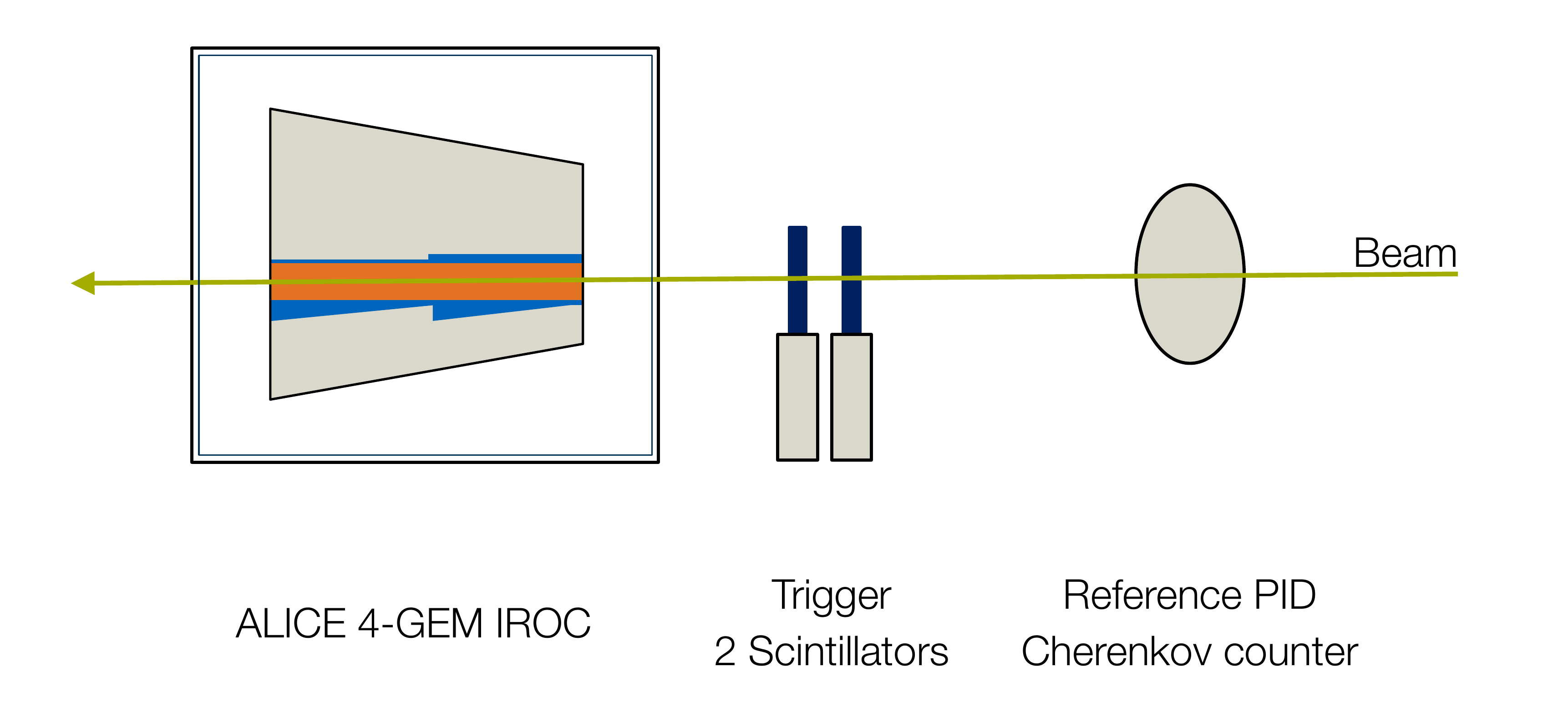}
\caption{Schematic top view of the setup at the T10 beam line (not to scale). A Cherenkov counter provides reference PID and two scintillators are used for beam definition and trigger. The trapezoidal 4-GEM IROC is enclosed in the test box with field cage. Moreover, a schematic representation of the readout region of the chamber (blue) and the fiducial region used in this analysis (orange) is displayed.}
\label{fig:setup}
\end{figure}

The beam line is equipped with several in-beam detectors which provide beam characterization and monitoring. At the end of the beam line, a threshold Cherenkov counter is mounted, which serves as a reference detector for particle identification. It is a cylindrical vessel filled with nitrogen at atmospheric pressure and equipped with a UV-sensitive photomultiplier tube, which detects the Cherenkov photons produced by traversing particles. 
For triggering, two scintillators are mounted in parallel with respect to each other at the exit window of the beam line. The trigger requires a coincidence of both modules. A schematic view of the setup is shown in \figref{fig:setup}.

\subsection{Readout electronics}
\label{subsec:readout}
The 4-GEM IROC prototype is equipped with 10 EUDET front-end cards (FECs) allowing for the readout of about 1200 channels. This configuration corresponds to a \SIrange{6}{7}{\centi\meter} wide corridor over the whole length of the IROC, as depicted by the blue area in \figref{fig:setup}. The FECs are based on the PCA16 \cite{PCA16}/ ALTRO chips \cite{ALTRO}. The ALTRO chip was originally designed for the ALICE TPC. In contrast to MWPCs, however, GEMs provide a negative-polarity input signal to the FEC. Therefore, a different preamplifier needs to be employed - the PCA16, specifically designed for Micro Pattern Gaseous Detectors. The sampling frequency is set to \SI{20}{\mega\hertz} at a peaking time of \SI{120}{\nano\second} and a conversion gain of \SI{12}{\milli\volt/\femto\coulomb}. 
The noise performance of the readout system corresponds to an RMS noise of about 600 $e^{-}$.

The similarity of the FECs to the ALICE FECs allows for the usage of the current TPC readout system \cite{NIMA_TPC}. The data are read out via the backplanes with two Readout Control Units (RCUs) \cite{Gutierrez:2005ffz} and transferred via optical links to a local data concentrator PC, which runs the ALICE data acquisition system DATE \cite{NIMA_TPC}. The data from the two scintillators, the Cherenkov counter and a sensor monitoring the ambient conditions are read out via CAMAC. The two data streams are synchronized based on an event tag.

The average DAQ rate is 300 events/spill with a spill length of \SI{0.5}{\second}, whereas the beam rate is of the order of 2000 particles/spill.

During the 7-day test beam campaign, the 4-GEM IROC was operated in \electron{} and $\pi^-$ beams at \SI{1}{\GeV/\clight}. A total of 3.6 million beam events were recorded, grouped into 142 data runs.
\section{Track reconstruction}
\label{sec:trackReco}
The front-end cards on the detector provide a digitized measurement of the amplitude on each pad at a certain time - a so-called \textit{digit}. In order to reconstruct the track of an incident particle from a set of digits, several reconstruction steps are performed.
The reconstruction is conducted with a dedicated framework, which partially relies on components of the AliRoot framework \cite{AliRoot}.

\subsection{Clustering}
The cluster finder starts by filling the digits sequentially into a pad-time matrix for each consecutive pad row and looking for amplitude maxima. A cluster is then created by investigating the amplitudes $A_{ij}$ stored in a $5\times5$ matrix with the maximum at its center. Pad hits are denoted by the index $i$, whereas time bin hits are denoted by $j$. 
The maximal charge of the cluster \Qmax{} = $A_{0,0}$ is given by the amplitude measured in the center of the matrix. For the computation of the total charge \Qtot{}, at first the bins adjacent to the central bin are investigated and the corresponding charge $A_{i,j}$ is summed up, i.e. 
\begin{equation}
\Qtot{} = \sum_{i=-1}^{1}\sum_{j=-1}^{1} A_{ij}.
\end{equation}
The charge of the outer bins, where $\lvert i \rvert\!=\! 2$ or $\lvert j \rvert\! =\! 2$, is only added to \Qtot{}, if the neighboring inner bin contains charge above threshold in order to reduce the contribution of electronic noise to the signal.

\subsection{Track fitting}
For the reconstruction of the test beam data, the pad and time coordinates of the found clusters are fitted independently by straight lines as a function of the pad row position.

The short drift length of the prototype leads to a significant amount of clusters with charge induced on a single pad only. 
The analysis of the residual distribution leads to a spatial resolution of about \SI{500}{\micro\meter} for the 4-GEM IROC.
However, this number is not representative for the performance of the final TPC in which the much longer drift length leads to sufficient charge sharing among several pads and hence to an improvement of the spatial resolution \cite{TDR_TPCU}.
\section{Data analysis}
\label{sec:analysis}
A number of selection criteria are applied to the recon\-struct\-ed tracks in order to prevent biases from background and edge effects. 
The usage of the Cherenkov counter as reference particle identification restricts the analysis to one-track events.
Only tracks with more than 32 associated clusters within a fiducial region around the center of the drift region are considered in the analysis in order to avoid any bias due to edge effects. Additional selection criteria are applied on the cluster drift time, as the cluster charge can be influenced by mismatched triggers in very low and high time bins. 
Moreover, tracks are rejected if more than 30\% of the associated clusters are outside the fiducial region in the pad direction (orange area in \figref{fig:setup}).

\subsection{Method of the truncated mean}
\label{subsec:truncMean}
The measured signal height is proportional to the energy deposit of the incident particle. As the typical mean free path of the particle is much smaller than the size of the readout pads, the charge liberated by several primary encounters is integrated by the readout.
For a sufficiently large number of measurements along a particle's trajectory, the distribution of the cluster charges $Q$ approaches the straggling function. Due to the structure of this distribution, its mean value \dEdxMean{} is not a good estimator of the energy deposit.
Therefore, the distribution of the cluster charges $Q$ is symmetrized by truncation. 
The truncated mean \dEdxTruncMean{} is here defined as the average of the 70\% lowest cluster charges associated to a track.
The resulting distribution is then fitted with a Gaussian and the relative resolution of \dEdxTruncMean{}, in the following referred to as \dEdx{} resolution, is defined as
\begin{equation}
\text{\dEdx{} resolution} = \frac{\sigma_{\dEdxTruncMean{}}}{\mu_{\dEdxTruncMean{}}},
\label{eq:dEdx}
\end{equation}
where $\sigma_{\dEdxTruncMean{}}$ and $\mu_{\dEdxTruncMean{}}$ are the standard deviation and mean value extracted from the fit, respectively.
The separation power of electrons and pions at the same momentum is given by
\begin{equation}
S_{\pi - e} = \frac{\lvert \mu_{ \langle\mathrm{d}E/\mathrm{d}x \rangle_{\mathrm{tr}, e^{-}}} - \mu_{ \langle\mathrm{d}E/\mathrm{d}x \rangle_{\mathrm{tr}, \pi^{-}} }  \rvert}{0.5 \cdot ( \sigma_{\langle\mathrm{d}E/\mathrm{d}x \rangle_{\mathrm{tr}, e^{-}}} + \sigma_{\langle\mathrm{d}E/\mathrm{d}x \rangle_{\mathrm{tr}, \pi^{-}} } ) }.
\label{eq:sep}
\end{equation}

\subsection{Corrections}
\label{subsec:corrections}

\begin{figure}
\centering
\includegraphics[width=\linewidth]{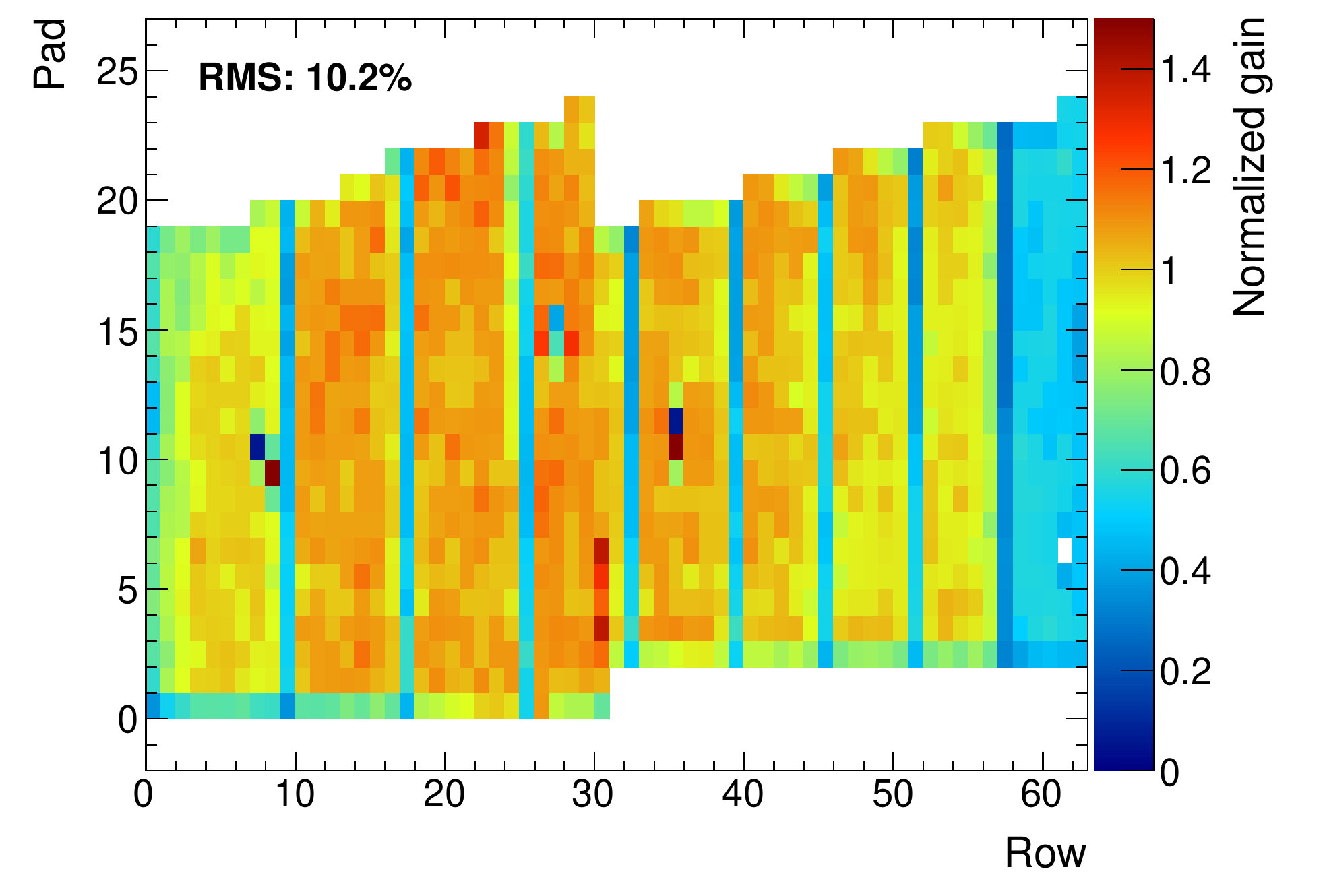}
\caption{Normalized gain map of the 4-GEM IROC prototype.}
\label{fig:normGain}
\end{figure}

Geometric imperfections of the GEM system as well as\break channel-by-channel variations of the front-end electronics may cause local variations of the gain. In order to minimize the impact of such variations, the normalized pad gain is extracted for each pad individually. The truncated energy loss \dEdxTruncMean{} is computed for each track and then the normalized cluster charge \Qtot{}/\dEdxTruncMean{} is monitored on each pad. 
This procedure assures that the determination of the normalized gain is independent of the absolute gain and the particle species. The normalized gain per pad is then determined by the most probable value of a Landau distribution fitted to the normalized cluster charge distribution on each pad separately.

Figure~\ref{fig:normGain} shows the resulting gain map of the 4-GEM IROC prototype. Clearly visible are pad rows with lower gain, which correspond to the GEMs' segment boundaries and the spacer grid geometrically overlapping the respective pad rows. 
A low-gain region for pad rows $>57$ is caused by a malfunctioning HV segment in the uppermost GEM. The distortions of the drift field, caused by the mismatching values of the field, last strip and the cover electrode potentials (see \secref{subsec:hvscheme}) lead to a slight loss of charge at the chamber borders which is visible in the first three rows of the gain map. For rows that are not affected by the aforementioned effects, the observed gain spread is about 10.2\% (RMS).

In order not to bias the \dEdx{} measurement, the charge of individual clusters is corrected by the normalized gain factor on a pad-by-pad basis. Rows with significantly reduced gain are not considered for the \dEdx{} computation. On average, 46 clusters (before truncation) are used for the calculation of the track d$E$/d$x$.

\subsection{Extraction of the \dEdx{} resolution}
As typically a few runs with the same HV configuration are taken, the significance of the measurement can be improved by merging the corresponding runs. 
Variations of environmental or running conditions, however, may alter the gain of the GEM stack and thus the performance of the detector. 
The variable under investigation (such as the separation power or the \dEdx{} resolution) is parametrized as a function of the gain, which varies with the environmental conditions. When these variations are found to be small, the behavior is parametrized using a polynomial of first order, while an exponential function is employed for larger variations.
This enables us to interpolate and to evaluate the variable in question at a gain of 2000.

\subsection{Systematic uncertainties}
The systematic uncertainty of the \dEdx{} resolution of individual runs is quantified by varying the selection criteria of the reference particle identification, on the time bins and the threshold for the acceptance cut by $\pm$20\%. The cut on the minimal number of clusters per track is kept unmodified in order not to bias the measurement due to the peculiar shape of the read out region. The resulting relative variation of the \dEdx{} resolution is found to be maximally about 0.3\%.

The systematic uncertainty introduced by the afore-\break mentioned interpolation procedure is investigated. The obtained results are compared to a simple average over different runs. In order not to bias this procedure, only runs with a 10\% variation around the desired gain are accepted. The relative deviation of the \dEdx{} resolution is then considered as the systematic uncertainty of the averaging procedure and found to be maximally $3-4$\%.
\section{Simulation}
\label{sec:sim}
In order to obtain a more solid understanding of the performance of the prototype, dedicated simulations are performed employing AliRoot \cite{AliRoot}, the framework for reconstruction, simulation and analysis in ALICE.

\subsection{AliRoot simulation}
\label{subsec:AliRootSim}
The TPC simulation in AliRoot is based on a modified version of GEANT3 \cite{GEANT3} and thoroughly described in \cite{TDR_TPC}. The distance between two successive ionizing collisions of the incident particle with the detector medium is randomly generated taking into account the expected average number of primary electrons per centimeter of track length.
At each step, a random energy loss according to $E^{-2.2}$ is assigned and the total number of ionization electrons is computed.
While drifting towards the readout anode, the diffusion and the attachment to electronegative gas impurities is taken into account for each electron independently.

At the GEM readout stage, the charge avalanche is then the convolution of single-electron avalanches, where random exponential gain fluctuations are taken into account. 
In order to assure a realistic treatment of the simulated data, pad-by-pad variations of the gain are simulated according to the measured performance of the 4-GEM IROC prototype (see Sec.~\ref{subsec:corrections}).
The pad response function is emulated taking into account the electron diffusion in the amplification region.
The signal shape is obtained by folding the electron avalanche with a semi-Gaussian shaping function.
The noise distribution is described by a Gaussian with an RMS of about 600 \electron, according to the performance of the readout electronics during the test beam. Finally, the signal is digitized by applying the given dynamic range of the electronics and zero suppression to the integrated signals created by the individual electrons.

The digitized signals are merged to clusters as discussed in Sec.~\ref{sec:trackReco}. The tracking procedure relies on the Kalman-filtering \cite{Kalman} approach and is thoroughly described in \cite{ALICE-Tracking}.

\subsection{Scaling of the electron detection efficiency}
\label{subsec:nPrimScaling}
The microscopic charge transport properties in a multi-GEM system, that lead to the characteristic dependence of \textit{IB} and \sigmaFe{} on the voltage settings, are not included in the simulation. Therefore, the deterioration of \sigmaFe{} is emulated to evaluate its impact on the global \dEdx{} resolution.
The deterioration is assumed to be caused by an effective detection loss of primary ionization during the charge transport through the GEM layers, i.e. that a fraction $x$ of the primary electrons is effectively lost. In order to compensate this loss of primary signal, the total gain in the simulation is increased by a factor $(1-x)^{-1}$ to maintain the same signal-over-noise ratio. 

\subsection{Characterization of the local energy resolution \sigmaFe{}}
A specific HV setting is characterized by its performance in terms of \textit{IB} and \sigmaFe{}. 
In order to be able to correlate the impact of the scaling of the electron detection efficiency to the performance of a specific HV configuration, the detector response to the \Fe{} source is investigated. To this end, 158 electrons are placed into the center of a pad row, corresponding to the ionization of a \SI{5.9}{keV} photon in the Ne-based gas mixture. 
As the amplification, digitization, clustering and tracking procedures may further influence the detector response, the generated signals are filtered through the same analysis chain as the actual data and simulations as described above.

\begin{figure}
\centering
\includegraphics[width=\linewidth]{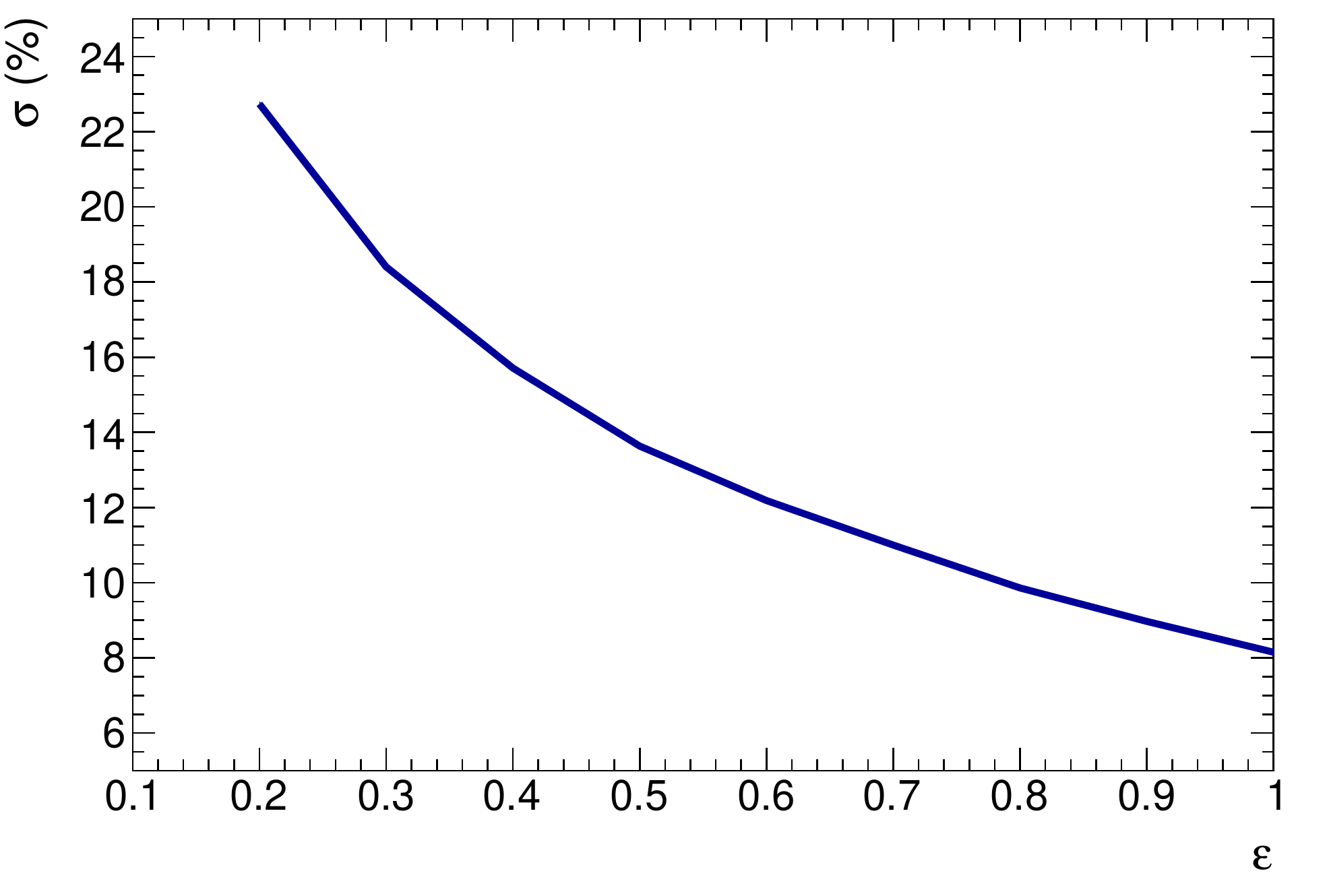}
\caption{Dependence of the local energy resolution $\sigma$ on the effective electron detection efficiency $\varepsilon=(1-x)$ obtained from simulated $^{55}$Fe signals. The uncertainties of the simulation is smaller than the line width.}
\label{fig:sigmaFe}
\end{figure}

The energy resolution is then extracted by fitting a Gaussian to the distribution of the total cluster charge $Q_{\mathrm{tot}}$. The degradation of the energy resolution as a function of the effective electron detection efficiency $\varepsilon = (1-x)$ is shown in \figref{fig:sigmaFe}. The observed resolution for $x=1$ is close to the expectation of $1/\sqrt{158} = 0.08$ for a detector with purely exponential gain fluctuations.

\section{Results}
\label{sec:results}

\subsection{Straggling functions}
\label{subsec:simulation}
\begin{figure}
\centering
\includegraphics[width=\linewidth]{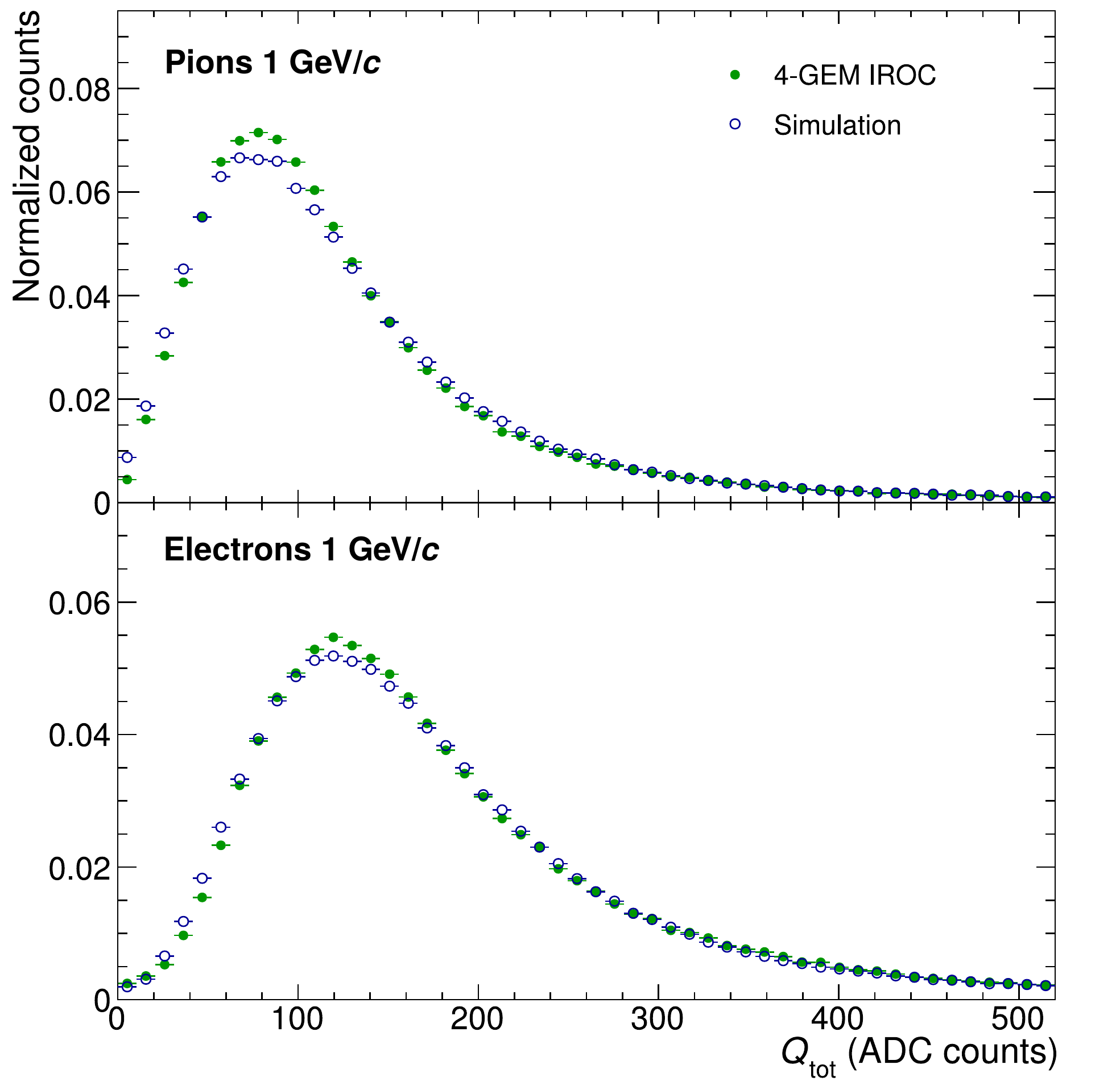}
\caption{The distribution of the total cluster charge \Qtot{} for pions (upper panel) and electrons (lower panel) at \SI{1}{\GeV/\clight} determined with experimental data measured with the 4-GEM IROC (closed symbols) compared to the simulation (open symbols).}
\label{fig:straggling}
\end{figure}

Detector effects, such as diffusion, gain variations, and the pad response, have been shown to significantly impact the\break \dEdx{} resolution \cite{PerformaceMWPCIROC, DetectorEffects} and hence need to be properly described in the simulation. Therefore, a comparison of the corresponding figures of merit of simulation and test beam data is mandatory to conclusively obtain a thorough understanding of the detector response.
In order to verify the performance of the simulation and to demonstrate that all physical processes involved in the signal formation are properly described, the first step is to compare the straggling functions of electrons and pions of the reconstructed data to the simulation. The straggling function is defined as the distribution of total charge \Qtot{} of all clusters originating from electron and pion tracks, respectively. The comparison, shown in \figref{fig:straggling}, demonstrates good agreement between data and simulation. 

\begin{figure}
\centering
\includegraphics[width=\linewidth]{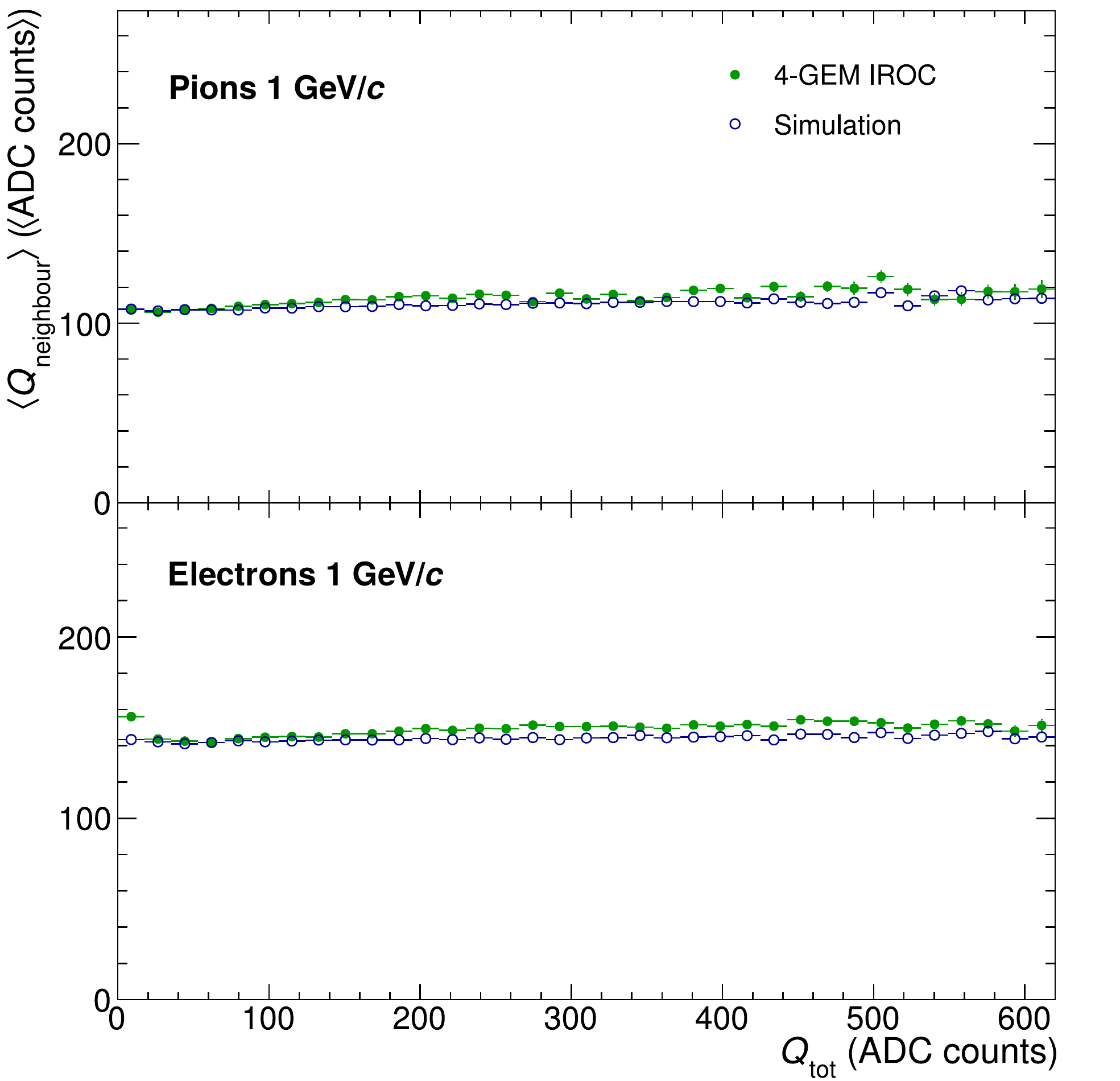}
\caption{The row couplings for pions (upper panel) and electrons (lower panel) at \SI{1}{\GeV/\clight} determined with experimental data measured with the 4-GEM IROC (closed symbols) compared to the simulation (open symbols).}
\label{fig:correlation}
\end{figure}

More information about the influence of the pad response function on the signal can be obtained by studying correlations between single rows. A non-projective contribution of the pad response function would cause charge leakage into adjacent pad rows, which would be visible in particular for large energy deposits. Therefore, the charge measured in one row is investigated as a function of the average charge in the neighboring row. 
In case the charges are uncorrelated, the distribution should be flat and the pad response function can be described as purely projective.
In \figref{fig:correlation}, a weak correlation is observed in data and simulation. The description by the simulation is reasonable, although a slightly stronger effect is observed in the data.

\subsection{d\textit{E}/d\textit{x} resolution and PID performance}
\label{subsec:dEdxresolution}

\begin{figure}[h]
\centering
\includegraphics[width=\linewidth]{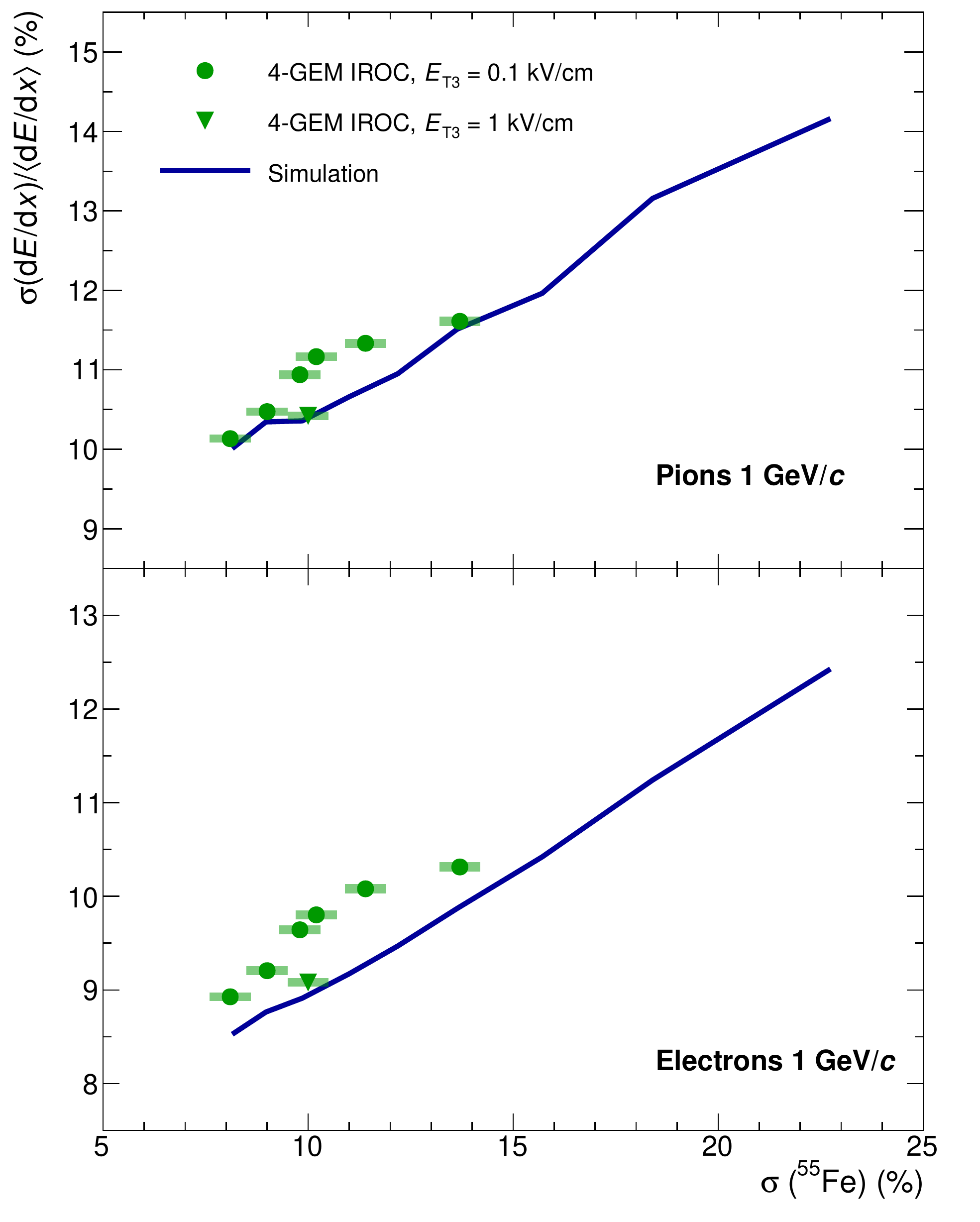}
\caption{\dEdx{} resolution measured with pions (upper panel) and electrons (lower panel) with different HV configurations corresponding to different values of \sigmaFe{} at a beam momentum of \SI{1}{\GeV/\clight}. The blue curve shows the result of a simulation. Statistical uncertainties are represented by lines, while systematical uncertainties are represented by boxes. The uncertainties of the simulation are smaller than the line width.}
\label{fig:dEdxElectron}
\end{figure}

The impact of the local energy resolution, expressed in terms of \sigmaFe{}, on the \dEdx{} resolution is studied by a systematic scan of the different HV configurations at a gain of 2000 as displayed in Tab.~\ref{tab:HVsettings2014}. The resulting \dEdx{} resolution of this study is shown in \figref{fig:dEdxElectron} for electrons and pions. As expected, the \dEdx{} resolution of the 4-GEM IROC decreases as the energy resolution degrades, even though the dependence is rather shallow and no sudden breakdown is observed. 
The resolution observed with on average 46 space points is well within the expectations of the current TPC \cite{NIMA_TPC}.
The simulations predict systematically better \dEdx{} resolution values for both electrons and pions. This points towards additional effects which seem to impact the energy resolution of the experimental setup, but may not yet be fully included into the simulation.
Notable is, however, that the general trend of the data is fairly well described by the simulations.

\begin{figure}[h]
\centering
\includegraphics[width=\linewidth]{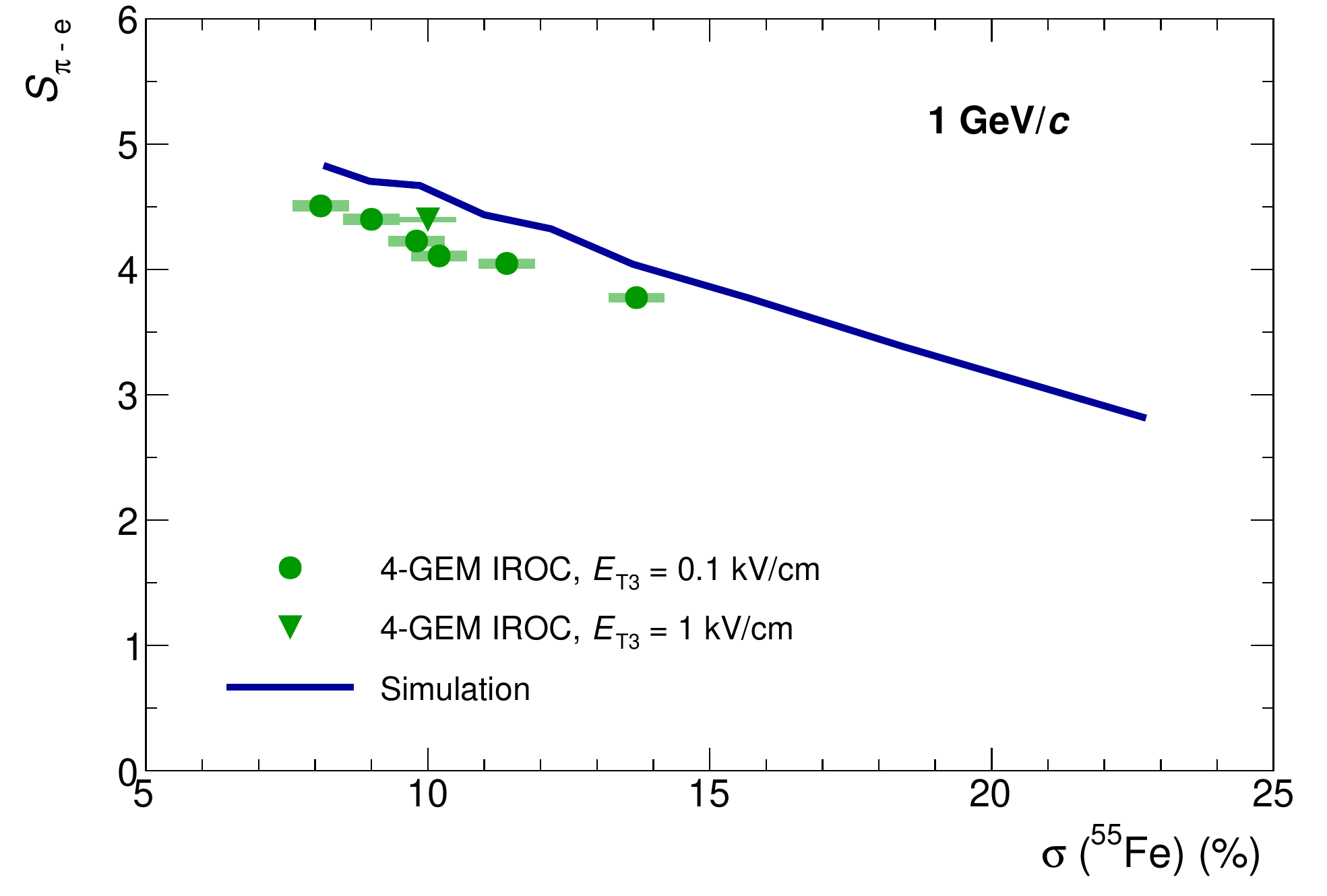}
\caption{Separation power of pions and electrons at \SI{1}{\GeV/\clight} as a function of \sigmaFe{}. The blue curve shows the result of a simulation. Statistical uncertainties are represented by lines, while systematical uncertainties are represented by boxes. The uncertainties of the simulation are smaller than the line width.}
\label{fig:separation}
\end{figure}

The corresponding separation power of the 4-GEM IROC is shown in \figref{fig:separation} as a function of \sigmaFe. As expected, the separation power decreases as \sigmaFe{} is degraded, even though the dependence is rather modest. The measured performance agrees fairly well with the simulation, although slightly better performance is predicted by the latter.

\begin{figure}[!h]
\centering
\includegraphics[width=\linewidth]{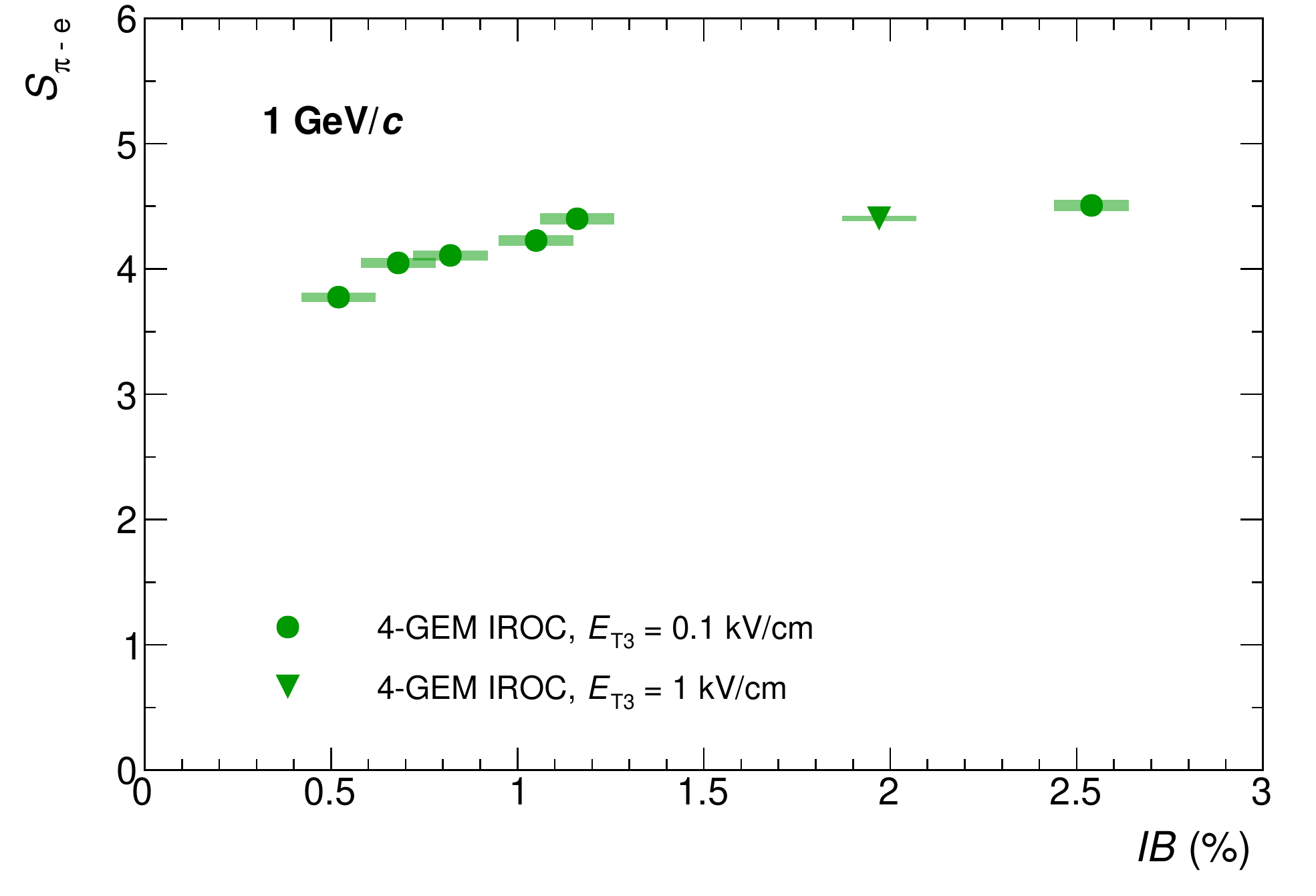}
\caption{Separation power of pions and electrons at \SI{1}{\GeV/\clight} as a function of the \textit{IB} for various HV settings. Statistical uncertainties are represented by lines, while systematical uncertainties are represented by boxes. }
\label{fig:ib}
\end{figure}

As anticipated, the PID performance decreases with decreasing value of \textit{IB} as displayed in \figref{fig:ib}. For larger values of \textit{IB}, only a slight improvement of the performance can be observed.
Notably, the HV configuration with the enhanced $E_{\mathrm{T3}}$ fits well in the overall trend of both the energy resolution and the \textit{IB}. This suggests that these two variables depend, to first order, only on the gain of the uppermost GEM foil in the stack.

For the operation of the TPC in Run 3 the detector will have to cope with a significant event pileup and space-charge distortions. 
Extensive studies \cite{TDR_TPCU, TDR_TPCU_add1} demonstrated that the \dEdx{} resolution slightly worsens with increasing occupancy from 5.5\% in isolated \pp{} events without pileup to about 7.5\% in central \PbPb{} at \SI{50}{\kHz}. This behavior is similar when using MWPC or GEM and is understood in terms of an increasing overlap of clusters.
The effect of space-charge distortions is corrected down to the intrinsic detector resolution employing sophisticated correction methods  \cite{TDR_TPCU, TDR_TPCU_add1}. Therefore, no additional degradation of the \dEdx{} resolution is expected.
\section{Summary}
\label{sec:summary}
The increased LHC luminosity envisaged for the Run 3 and beyond implies significant upgrades of the ALICE TPC, as the gating grid of the current MWPC-based readout chambers imposes unacceptable rate limitations. 
The \dEdx{} resolution is a crucial detector parameter for the particle identification via measurement of the specific energy loss. As the ALICE TPC is the main device for PID in ALICE, it is therefore of particular importance to ensure its excellent performance is retained after the upgrade. This work presents the result of a comprehensive \dEdx{} resolution study conducted with a 4-GEM IROC prototype for the upgrade of the ALICE TPC.

Local variations of the electron amplification in the GEM stack are quantified with tracks from the test beam to about 10\% and thus found to be well within the requirements.
The \dEdx{} resolution and hence the PID performance of the prototype is compatible with the requirements of the ALICE upgrade physics programme. 
The results are compared to the outcome of a detailed simulation, in which the local energy resolution is degraded by decreasing the electron detection efficiency of the readout system. The experimental data are reasonably well described by the model.

We conclude that the 4-GEM stack fully provides the performance required for the upgrade of the ALICE TPC, which endorses its choice as the baseline solution.

\newenvironment{acknowledgement}{\relax}{\relax}
\begin{acknowledgement}
\section*{Acknowledgments}
The ALICE TPC Collaboration would like to thank the \break LCTPC Collaboration for borrowing the readout electronics for the PS test beam and the RD51 Collaboration for useful discussions.

The ALICE TPC Collaboration would like to thank the\break CERN accelerator teams for the
outstanding performance of the Proton Synchrotron during the test beam.

The ALICE TPC Collaboration acknowledges the following funding agencies for their support in the TPC Upgrade:
Funda\c{c}\~ao de Amparo \`a Pesquisa do Estado de S\~ao Paulo\break (FAPESP), Brasil;
Ministry of Science and Education, Croatia;
The Danish Council for Independent Research | Natural Sciences, the Carlsberg Foundation and Danish National Research Foundation (DNRF), Denmark;
Helsinki Institute of Physics (HIP) and Academy of Finland, Finland;
Bundesministerium f\"{u}r Bildung, Wissenschaft, Forschung und Technologie (BMBF), 
GSI Helmholtzzentrum f\"{u}r Schwerionenforschung\break GmbH, 
DFG Cluster of Excellence "Origin and Structure of the Universe", 
The Helmholtz International Center for FAIR (HIC for FAIR)
and the ExtreMe Matter Institute EMMI at the GSI Helmholtzzentrum f\"{u}r Schwerionenforschung, Germany;
National Research, Development and Innovation Office, Hungary;
Department of Atomic Energy Government of India (DAE)
and Department of Science and Technology, Government of India (DST), India;
Nagasaki Institute of Applied Science (IIST)
and the University of Tokyo, Japan;
Fondo de Cooperaci\'{o}n Internacional en Ciencia y Technolog\'{i}­a (FONCICYT), Mexico;
The Research Council of Norway, Norway;
Ministry of Education, Science, Research and Sport of the Slovak Republic, Slovakia;
Swedish Research Council (VR), Sweden;
United States Department of Energy, Office of Nuclear Physics (DOE NP), USA.

\end{acknowledgement}
\bibliographystyle{elsarticle-num} 
\bibliography{BibtexDatabase.bbl}
\end{document}